\documentclass[%
reprint,
 amsmath,amssymb,
 aps,
prd,
tightenlines
]{revtex4-1}

\usepackage{color}
\usepackage{multirow}
\usepackage{overpic}
\usepackage{colortbl}
\usepackage{graphicx}
\usepackage{float}
\usepackage{array}
\usepackage{makecell}
\usepackage{subfigure}
\usepackage{graphicx}
\usepackage{dcolumn}
\usepackage{bm}
\usepackage{xcolor}
\usepackage{amsmath}
\usepackage[T1]{fontenc}
\definecolor{aa}{RGB}{0,0,139}

\newcommand{\jpsi}{J/\psi}

\newcommand{\ppjpsi}{\pi^+\pi^-\jpsi}

\newcommand{\lum}{{\cal L}}
\newcommand{\eff}{\varepsilon}
\newcommand{\BR}{{\cal B}}

\newcommand{\xx}{X(3872)}
\newcommand{\xs}{X(3700)}

\newcommand{\etap}{\eta^{\prime}}

\newcommand{\psip}{\psi(2S)}
\newcommand{\pspp}{\psi(3770)}

\newcommand{\EE}{e^+e^-}
\newcommand{\MM}{\mu^+\mu^-}
\newcommand{\LL}{\ell^+\ell^-}
\newcommand{\pp}{\pi^+\pi^-}

\newcommand{\ddbar}{D^{0}{\bar{D}}^{0}}
\newcommand{\ccb}{c\bar{c}}

\newcommand{\gampipi}{$\gamma\pi^{+}\pi^{-}$}

\usepackage{multirow}
\usepackage{enumerate}
\usepackage{amsmath}
\begin{document}


\title{\boldmath Search for a scalar partner of the $\xx$ via
$\pspp$ decays into $\gamma\eta\etap$ and $\gamma\ppjpsi$}

\author{
M.~Ablikim$^{1}$, M.~N.~Achasov$^{13,b}$, P.~Adlarson$^{75}$, X.~C.~Ai$^{81}$, R.~Aliberti$^{36}$, A.~Amoroso$^{74A,74C}$, M.~R.~An$^{40}$, Q.~An$^{71,58}$, Y.~Bai$^{57}$, O.~Bakina$^{37}$
% ...
\\
\vspace{0.2cm}
(BESIII Collaboration)\\
\vspace{0.2cm} {\it
$^{1}$ Institute of High Energy Physics, Beijing 100049, People's Republic of China\\
% ...
$^{13}$ G.I. Budker Institute of Nuclear Physics SB RAS (BINP), Novosibirsk 630090, Russia\\
% ...
$^{36}$ Johannes Gutenberg University of Mainz, Johann-Joachim-Becher-Weg 45, D-55099 Mainz, Germany\\
$^{37}$ Joint Institute for Nuclear Research, 141980 Dubna, Moscow region, Russia\\
% ...
$^{40}$ Liaoning Normal University, Dalian 116029, People's Republic of China\\
% ...
$^{57}$ Southeast University, Nanjing 211100, People's Republic of China\\
$^{58}$ State Key Laboratory of Particle Detection and Electronics, Beijing 100049, Hefei 230026, People's Republic of China\\
% ...
$^{71}$ University of Science and Technology of China, Hefei 230026, People's Republic of China\\
% ...
$^{74}$ University of Turin and INFN, (A)University of Turin, I-10125, Turin, Italy; (B)University of Eastern Piedmont, I-15121, Alessandria, Italy; (C)INFN, I-10125, Turin, Italy\\
$^{75}$ Uppsala University, Box 516, SE-75120 Uppsala, Sweden\\
% ...
$^{81}$ Zhengzhou University, Zhengzhou 450001, People's Republic of China\\
\vspace{0.2cm}
% ...
$^{b}$ Also at the Novosibirsk State University, Novosibirsk, 630090, Russia\\
% ...
}}

\date{\today}

\begin{abstract}

Using a data sample corresponding to an integrated luminosity of 2.93 fb$^{-1}$ collected at a center-of-mass energy
of 3.773~GeV with the BESIII detector at the BEPCII collider, we search for
a scalar partner of the $\xx$, denoted as $\xs$, via $\pspp\to \gamma\eta\etap$ and $\gamma\ppjpsi$ processes.
No significant signals are observed and the upper limits of the product
branching fractions
  $\BR(\pspp\to\gamma \xs)\cdot \BR(\xs\to \eta\etap)$
and
  $\BR(\pspp\to\gamma \xs)\cdot \BR(\xs\to \ppjpsi)$
are determined at the 90\% confidence level, for the narrow $\xs$ with a
mass ranging from 3710 to 3740~MeV/$c^2$, which are from 0.9 to 1.9 $(\times 10^{-5})$ and 0.9 to 3.4 $(\times 10^{-5})$, respectively.

\end{abstract}

\pacs{Valid PACS appear here}

\maketitle

\section{Introduction}
Since 2003, a number of resonances that decay to final states with a pair of $c\bar{c}$ quarks have been discovered~\cite{Bevan:2014iga,LHCb:2019quf, LHCb:2013kgk,Yuan:2021wpg}. Whereas some of these states have properties well expected for conventional $\ccb$ mesons~\cite{Deng:2016stx}, and others have properties that do not match those of any $\ccb$ mesons and can be described by configurations beyond quark model, such as multi-quark or hadronic-molecule ~\cite{Brambilla:2010cs,Brodsky:2014xia,Chen:2016qju,Olsen:2017bmm,Guo:2017jvc,Brambilla:2019esw}.
Among these states, the $\xx$ ($I^{G}(J^{PC})=0^{+}(1^{++})$), discovered at the Belle experiment in 2003~\cite{Belle:2003nnu},
attracted lots of attention due to the fact that it is very close to the
$D^{*0}\bar{D}^0$ threshold (Q = $m_{\xx}-m_{D^{*0}}-m_{\bar{D}^{0}} = (-0.04\pm0.09)$ MeV/$c^{2}$)
and very narrow ($\Gamma$ = $(1.19\pm0.21)$ MeV)~\cite{int22}. As a good candidate for the
$D^{*0}\bar{D}^0$ molecule, the $\xx$ stimulated many studies of similar
states at other meson pair thresholds, such as $X_{2}(4013)$ (expected to be an $S$-wave $D^{*}\bar{D}^{*}$ bound state)~\cite{Nieves:2012tt}.

In the hadronic-molecular picture, the $\xx$
can have some degenerated spin partners with quantum numbers $0^{++}$, $1^{++}$ and $2^{++}$
in the heavy quark limit~\cite{int13, int14, int15}.
As a good candidate for the $0^{++}$ state, the $\xs$ ($I^{G}(J^{PC})=0^{+}(0^{++})$), a $D\bar{D}$ molecule,
is predicted with a mass around 3720~MeV/$c^{2}$ via heavy quark-spin-symmetry
arguments~\cite{ int14, int15}.
The discovery of such a $0^{++}$ charmonium-like state would provide a strong support
for the interpretation of the $\xx$ as a hadronic-molecule dominant state.
In addition, the $\xs$ is predicted by lattice quantum chromodynamics in the study of the coupled-channel $D\bar{D} - D_{s}\bar{D_{s}}$
scattering~\cite{int12}, and it is found as a shallow $D\bar{D}$ bound state just below
the $D\bar{D}$ threshold.
In chiral unitary theory, the $\xs$ has already appeared as a pole in the T-matrix in the study of dynamical generation of resonances,
and its structure is mainly a $D\bar{D}$ quasibound state~\cite{int11}.

Experimentally, some hints of the existence of $\xs$ have been reported. A bump in the $D\bar{D}$ spectrum close to the $D\bar{D}$ threshold observed by the Belle Collaboration in the $\EE\to J/\psi D\bar{D}$ reaction was better interpreted in terms of the bound state below threshold, with $M_{\xs} = 3723$~MeV/$c^2$~\cite{Gamermann:2007mu} and $M_{\xs} = 3706$~MeV/$c^2$~\cite{Wang:2019evy} than with a resonance $X(3880)$ as suggested in Ref.~\cite{int19}.
Besides, a hint of enhancement just above the $D\bar{D}$ threshold was both seen by the BaBar and Belle Collaborations in the reaction $\gamma\gamma\to D\bar{D}$~\cite{int20, Belle:2005rte}.
By fitting to the $D\bar{D}$ invariant mass distributions measured by the Belle and BaBar Collaborations, taking into account the $S$-wave $D\bar{D}$ final state interaction, the existence of $\xs$ has been further investigated in Ref.~\cite{Wang:2020elp}.

Although the $\xs$ has been predicted by many different models,
it has not been confirmed by any experiments yet. Furthermore, as a $D\bar{D}$ bound state,
the $\xs$ can not be generated by the one-pion exchange interaction between the $D$ mesons due to the $P$ parity conservation in the strong interaction.
Various strategies for the experimental search for
the $\xs$ in exclusive decays have been proposed, e.g. $\pspp\to\gamma\eta\etap$~\cite{int21},
$\pspp\to\gamma \ddbar$~\cite{Dai:2020yfu} and $B^{0(+)}\to\ddbar K^{0(+)}$~\cite{Dai:2015bcc}.
The dominant decay of the $\xs$ is to $\eta\etap$,
with the partial decay width $\Gamma _{\pspp\to\gamma \xs,~\xs\to\eta\etap}$ = 0.293~keV according to the prediction based on the chiral unitary theory~\cite{int21}.
Since the total width $\Gamma _{\pspp} = (27.2\pm1.0)$~MeV~\cite{int22}, we get
$\BR(\pspp\to\gamma\xs) \cdot \BR(\xs\to\eta\etap) = \Gamma_{\pspp\to\gamma\xs,~\xs\to\eta\etap}/\Gamma _{\pspp} \simeq 1.08\times10^{-5}$.
Therefore, the process $\pspp\to\gamma\eta\etap$ is studied to test the theoretical prediction.
We also search for the $\xs$ via $\xs\to\ppjpsi$, since the $\xs$ decay might
have a large isospin violation similar to $\xx\to\ppjpsi$~\cite{int24}. It will help to test the heavy quark-spin-symmetry and deepen our understanding of the nature of both the $\xx$ and $\xs$.

In this article, we report the studies of the two processes $\pspp\to\gamma\eta\etap$
and $\gamma\ppjpsi$, with $\eta\to\gamma\gamma$, $\etap\to\gamma\pp$ and
$\jpsi\to\LL (\ell = e,~\mu)$, using the data sample corresponding to an integrated luminosity of 2.93~fb$^{-1}$ collected with the BESIII detector in the years 2010 and 2011, at center-of-mass (c.m.) energy  $\sqrt s=3.773$ GeV~\cite{data1, data2}.

\section{BESIII Detector and Monte Carlo Simulation}\label{sec:dec}
The BESIII detector~\cite{dec1} records symmetric $e^+e^-$ collisions
provided by the BEPCII storage ring~\cite{dec2}
in the c.m. energy range from 2.0 to 4.95~GeV, with a peak luminosity of $1 \times 10^{33}~\text{cm}^{-2}\text{s}^{-1}$
achieved at $\sqrt{s} = 3.77~\text{GeV}$.
The cylindrical core of BESIII detector consists of a helium-based multilayer drift chamber (MDC),
a plastic scintillator time-of-flight system (TOF), and a CsI (Tl) electromagnetic calorimeter (EMC),
which are all enclosed in a superconducting solenoidal magnet providing a 1.0~T magnetic field~\cite{detvis}.
The solenoid is supported by an octagonal flux-return yoke with resistive plate counter muon
identifier modules interleaved with steel. The acceptance of charged particles and photons
is 93\% over 4$\pi$ solid angle. The charged particle momentum resolution at 1 GeV/$c$ is 0.5\%,
and the $dE/dx$ resolution is 6\% for electrons from Bhabha scattering.
The EMC measures photon energies with a resolution of 2.5\% (5\%) at 1~GeV in the barral (end cap) region.
The time resolution of the TOF barrel part is 68~ps, while that of the end cap part is 110~ps.

The optimization of event selection criteria and efficiency determination are based on
Monte Carlo (MC) simulations. The {\sc geant4}-based~\cite{data3} simulation software, BESIII Object Oriented Simulation Tool ({\sc boost})~\cite{data4}, includes the geometric description of the BESIII detectors.
The simulation models the beam energy spread and initial state radiation (ISR) in the
$\EE$ annihilations with the generator {\sc kkmc}~\cite{data5, data51}.
Samples of MC simulated events for the signal decay $\pspp\to \gamma\xs$ are generated with the
P2GC0 model in {\sc{evtgen}}~\cite{data7, data71}. The subsequent decays $\xs\to \eta\etap$
and $\rho^0 \jpsi$ are generated uniformly in phase space.
$\rho^0\to\pi^+\pi^-$ is generated with VSS model in {\sc{evtgen}}~\cite{data7, data71}.
Final state radiation from
charged final state particles is incorporated using {\sc photos} package~\cite{data6}.
The mass of the $\xs$ is assumed to be 3710, 3715, 3720, 3725, 3730, 3733, 3735, and
3740~MeV/$c^2$ and the width is set to 1.0~MeV~\cite{int11} in the simulation.

To estimate the possible background, we use the inclusive MC sample generated at $\sqrt{s}=3.773$~GeV.
The inclusive MC sample includes the production of $D\bar{D}$
pairs (including quantum coherence for the neutral $D$ channels),
the non-$D\bar{D}$ decays of the $\psi(3770)$, the ISR
production of the $J/\psi$ and $\psi(2S)$ states , and the
continuum processes incorporated in {\sc kkmc}~\cite{data5}.
The statistics of $D\bar{D}$ pairs is 10.8 times that of data, the statistics of non-$D\bar{D}$, ISR $J/\psi$ and ISR $\psi(2S)$ are 10.2 times that of data, and the statistics of continuous process is 7.4 times that of data.
The known decay modes are generated with {\sc evtgen} with branching
fractions being set to the world average values~\cite{int22} and the remaining events are generated with {\sc lundcharm}~\cite{data9, data91} while other
hadronic events are generated with {\sc pythia}~\cite{Sjostrand:2014zea}.

\section{Event Selection}
Charged tracks detected in the MDC are required to be within a polar angle ($\theta$) range of $|\rm{cos\theta}|<0.93$, where $\theta$ is defined with respect to the $z$-axis, which is the symmetric axis of the MDC.
For charged tracks, the distance of closest approach to the interaction point
must be less than 10~cm along the $z$-axis, $|V_{z}|$,
and less than 1~cm in the transverse plane, $|V_{xy}|$.

Photon candidates are identified using showers in the EMC.  The deposited energy of each shower must be more than 25~MeV in the barrel region ($|\cos \theta|< 0.80$) and more than 50~MeV in the end cap region ($0.86 <|\cos \theta|< 0.92$).
To significantly reduce showers that originate from charged tracks, the angle subtended by the EMC shower and the position of the closest charged track at the EMC must be larger than 10$^{o}$ as measured from the interaction point. To suppress electronic noise and showers unrelated to the event, the difference between the EMC time and the event start time is required to be within [0, 700]~ns.

For the $\pspp\to \gamma\eta\etap$, $\eta\to \gamma\gamma$ and $\etap\to \gamma\pi^+\pi^-$ channel,
we choose $\etap\to\gamma\pi^+\pi^-$ with the largest branching fraction (taking into account the branching fraction of the intermediate resonance decays).
Event candidates are required to have exactly two charged tracks with zero net-charge and at least four photons.
The two charged tracks are assumed as pions without using particle identification information.
A vertex fit is performed on the two charged tracks to make sure that they originate from
the same vertex. In order to select $\eta$ candidate, we do pairwise combinations for
all the photons, and require the sum of the two photon energies to be larger than 1.65~GeV
and their invariant mass satisfies $0.51 < M(\gamma\gamma) < 0.57$~GeV/$c^2$.
To improve momentum and energy resolutions and suppress background, a five-constraint (5C) kinematic fit, which constrains
the sum of four momentum of the final-state particles to the initial momentum of the colliding
beams and the invariant mass of the $\eta$ candidate to the $\eta$ world average
mass~\cite{int22},
is performed to an event with the hypothesis $\pspp\to \gamma\gamma\eta\pp$, and $\chi^2_{\rm 5C}<18$ is required.
If there are more than one combination due to multiple $\eta$
candidates or photons, the combination with the minimum $\chi^2_{\rm 5C}$ is retained.

The $\gamma\pp$ combination with the minimum $|M(\gamma\pp)-m_{\etap}|$ is chosen
as the $\etap$ candidate and events with $|M(\gamma\pp)-m_{\etap}|<0.009$~GeV/$c^2$
are selected for further analysis.
The $\pp$ invariant mass is required to be in the $\rho^0$ mass region,
$0.6< M(\pp) < 0.8$~GeV/$c^2$. The asymmetric mass window for $\rho^0$ is chosen following Ref.~\cite{BESIII:2017kyd}.
To suppress background containing a $\pi^0$,
events with $|M(\gamma\gamma)-m_{\pi^0}|<0.015$~GeV/$c^2$
are rejected, where $M(\gamma\gamma)$ is the invariant mass of any photon pairs
passed the kinematic fit and $m_{\pi^0}$ is the nominal mass of the $\pi^{0}$ from the PDG~\cite{int22}.

For the $\pspp\to \gamma\ppjpsi$, $\jpsi\to \l^{+}\l^{-} (\EE$ or $\MM$) channel, events
with four good charged tracks with zero net-charge and at least one photon candidate are selected.
Since the $\pi^{\pm}$ in the final state and $\ell^{\pm}$ from $\jpsi$ decay are kinematically
well separated, charged tracks with momenta larger than 1.2~GeV/$c$ in the laboratory frame
are assumed to be $\ell^{\pm}$ and those with momenta less than 0.5~GeV/$c$
are assumed to be $\pi^{\pm}$.
The energy deposition of charged track in the EMC is used to separate $e$ and $\mu$.
For $\mu$ candidate, the deposited energy in the EMC is required to be less than 0.5~GeV,
while it is required to be larger than 1.1~GeV for $e$.
To improve the momentum and energy resolutions and reduce background, a 4C kinematic fit, which constrains
the sum of four momentum of the final-state particles to the initial momentum of the colliding beams, is
applied to an event with the hypothesis $\pspp\to \gamma\pp \LL$, and $\chi_{\rm 4C}^2<36$ (34) is required for $\EE$ ($\MM$) mode. If there is more than one photon, the combination with the least $\chi_{\rm 4C}^2$ is chosen.
The $\jpsi$ signal mass window of $3.087<M(\LL)<3.120$~GeV/$c^2$, and the
sidebands of $3.05<M(\LL)<3.07$~GeV/$c^2$ and $3.13<M(\LL)<3.15$~GeV/$c^2$, are further required.

In order to suppress the background contributions from low momentum electrons mis-identified as pions,
pion candidates are identified using the $dE/dx$ information recorded in the MDC.
A discriminator $\chi_{\pi^{\pm}} = (\mu_{\rm m} - \mu_{\rm exp})/\sigma_{\rm m}$ is defined
by combining the measured $dE/dx$ value ($\mu_{\rm m}$), the measurement uncertainty
($\sigma_{\rm m}$) and the expected value under a pion hypothesis ($\mu_{\rm exp}$).
The conditions $\chi_{\pi^+}<3.$0 and $\chi_{\pi^-}<$3.0 are applied to provide an optimal balance between efficiency loss and background rejection power.
To further reject radiative Bhabha ($\gamma \EE$) background associated with photon conversion,
the cosine of the opening angle of the $\pp$ candidates in the laboratory frame is required to be less than 0.95
in $\jpsi\to \EE$ mode. This requirement removes almost all the photon conversion background events with an efficiency loss
less than 1\%.
The remaining background events mainly come from $\EE\to\gamma_{\rm ISR}\psip$ (with $\psip\to\ppjpsi$),
which has the same final state as the signal mode. To suppress such background,
$|\cos\theta_{\gamma}|$ is required to be less than 0.75, where $\theta_{\gamma}$ is the polar
angle of the radiative photon in the laboratory frame.

\section{SIGNAL EXTRACTION}
\begin{figure*}[htbp]
  \centering
  \subfigure[]{
  \label{Fig.sub.1}
  \includegraphics[width=0.45\textwidth]{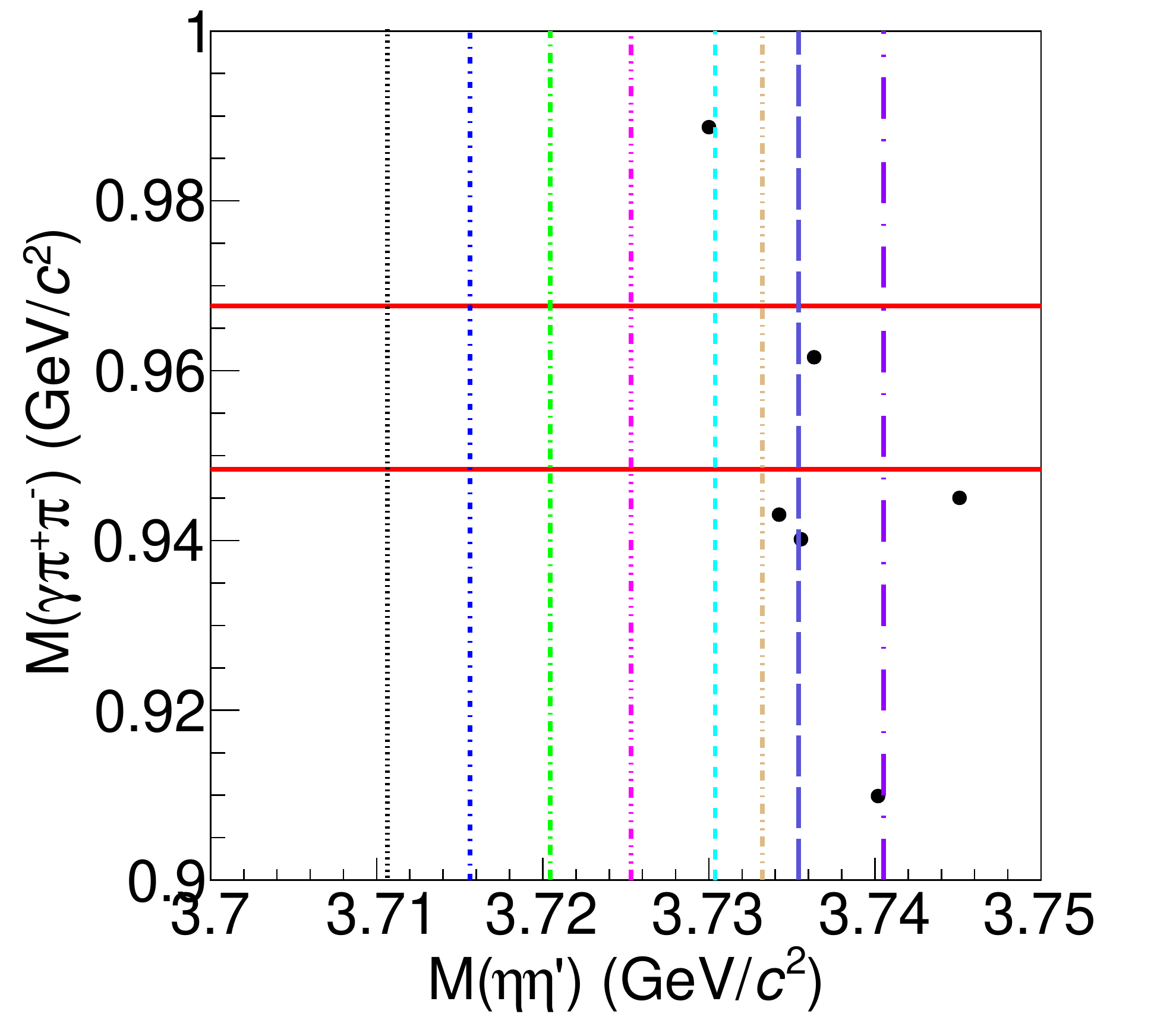}}
  \subfigure[]{
  \label{Fig.sub.2}
  \includegraphics[width=0.45\textwidth]{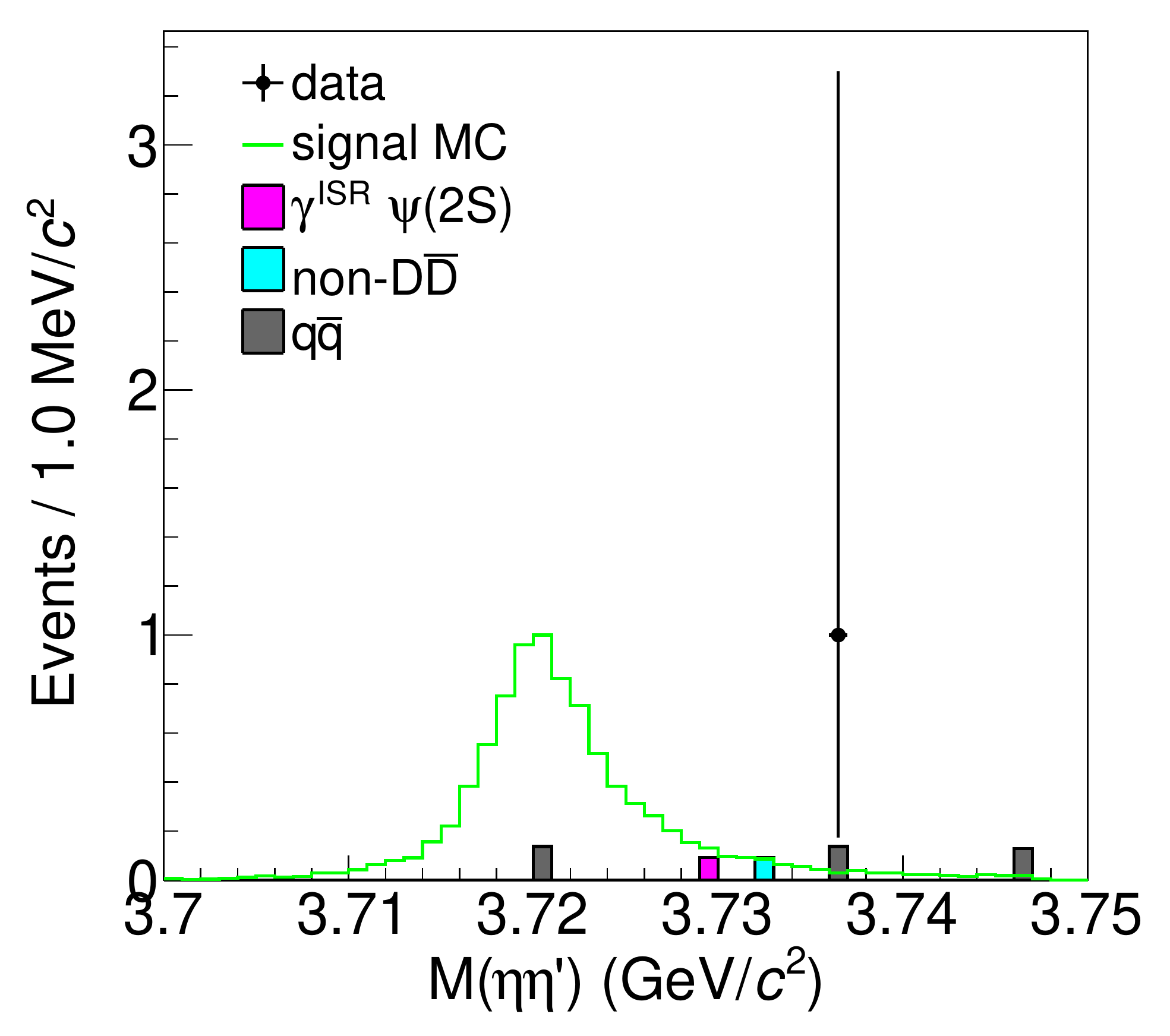}}
\caption{The two-dimensional distribution of $M(\gamma\pp)$ versus $M(\eta\etap)$ from the data sample (a)
and the distribution of $M(\eta\etap)$ after $\etap$ selection from the data, inclusive MC,
and signal MC samples (b). In (a), the red solid lines mark the signal
region of $\eta'$ and other colored dashed lines mark the nominal $\xs$ mass with different mass assumptions (detailed information can be found in Sec.~\ref{sec:dec}); in (b), the black dots with error bars represent the data sample,
the green solid line represents the signal MC sample with $\xs$ mass of 3720 MeV/$c^{2}$ ,
and other colored shaded histograms represent the inclusive MC samples.
The number of events in the inclusive MC sample is normalized according to the integrated
luminosity of the data sample. The maximum bin content is set to one for the signal MC sample.}
\label{pic:Metaetap-etap-data}
\end{figure*}

For the $\pspp\to \gamma \eta\etap$ mode,
Fig.~\ref{pic:Metaetap-etap-data}(a) shows the two-dimensional distribution of
$M(\gamma\pp)$ versus $M(\eta\etap)$ from the data sample after imposing all the requirements
mentioned above, and Fig.~\ref{pic:Metaetap-etap-data}(b) shows the $\eta\etap$
invariant mass distributions from the data, inclusive MC and signal MC samples
after further tagging an $\etap$ candidate by the criterion, $|M(\gamma\pp)-m_{\etap}|< 0.009 $~GeV/$c^2$.
Only one candidate event survives the selection between 3.70 and 3.75~GeV/$c^2$, and no significant $X(3700)$ signals are observed in this channel.

A detailed study of the inclusive MC sample indicates that there are only a few events
survived and distributed randomly in this mass region. We used a ``cut and count'' method to
extract the number of signal events for different $\xs$ masses.
The signal events are selected with both $M(\eta\etap$) and $M(\gamma\pp$)
within mass windows of twice the mass resolutions ($2\sigma$) around their mean values.
Here the mean values stand for the nominal masses of $\xs$ and $\etap$.
The mass resolutions are determined from the simulation.

\begin{figure*}[htbp]
  \centering
  \includegraphics[width=0.9\textwidth]{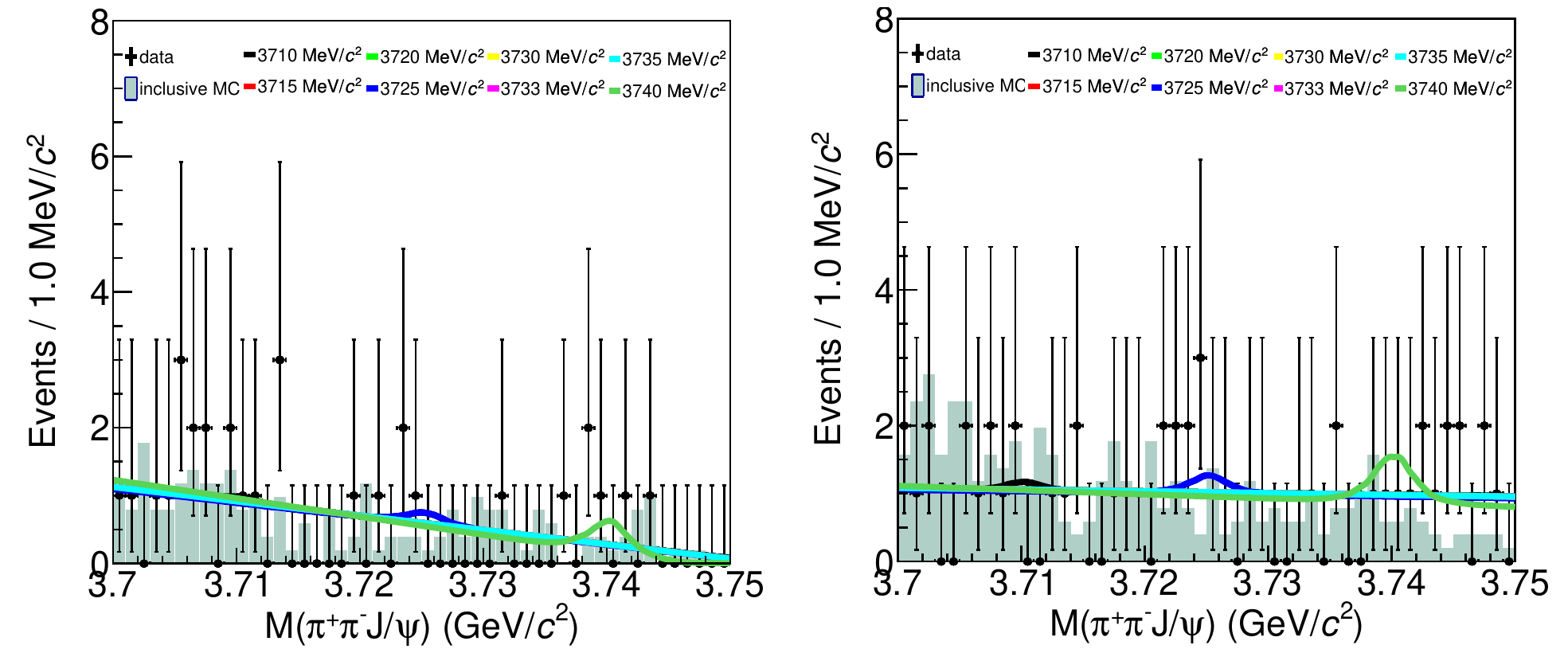}
\caption{The simultaneous unbinned maximum likelihood fit on the $M(\ppjpsi)$ distributions
for $\EE$ mode (left) and $\MM$ mode (right).
Black dots with error bars represent the data sample,
and other colored shaded histograms represent the inclusive MC sample.
The solid lines with different color represent the total fit results for
different masses of the $\xs$.
The number of events in the inclusive MC sample is
normalized according to the integrated luminosity of the data sample.}
\label{pic:Mx}
\end{figure*}

For the $\pspp\to \gamma\ppjpsi$ mode, the plots in Fig.~\ref{pic:Mx}
show the $\ppjpsi$ invariant mass distributions from the data and inclusive MC sample after imposing all the requirements mentioned above.
Here $M(\ppjpsi) = M(\pp\LL) - M(\LL) + m_{\jpsi}$ is used to reduce the resolution effect of the lepton pairs~\cite{sys10}, and $m_{\jpsi}$ is the nominal mass of the $\jpsi$ meson~\cite{int22}. No significant $\xs$ signals are observed in this
channel either. The surviving events are dominated by backgrounds from
$\EE\to \gamma_{\rm ISR}\psip$ (with $\psip\to \ppjpsi$).

To extract the signal yields, we perform a simultaneous unbinned maximum likelihood fit
on the $M(\ppjpsi)$ distributions for both $\jpsi\to \EE$ and $\jpsi\to \MM$ modes.
In the simultaneous fit, the signal is described by the line shape of the $\xs$
from MC simulation, and the background is described by a linear function. The fit results are shown in Fig.~\ref{pic:Mx}. Since the mass of $\xs$ is not determined yet, various fits are performed under different $\xs$ mass assumptions, and no obvious the $\xs\to\ppjpsi$ signals are observed from the various fits.

Since there are no significant $\xs$ signals, we set the upper limit of the
product branching fractions for $\pspp\to\gamma \xs$, $\xs\to \eta\etap$ and $\xs\to\ppjpsi$.
For $\xs\to \eta\etap$ and $\xs\to\ppjpsi$, we have not observed obvious $\xs$ signals from the $\pspp$ resonance data and continuum data, so it is difficult to consider the contribution from the continuous process and the interference between the $\psi(3770)$ and the continuum amplitudes.
We assume that all the contribution is from $\pspp$ decay and the interference between the $\psi(3770)$ and the continuum amplitudes~\cite{guo} is ignored.
The upper limit of the branching fraction of
$\pspp\to \gamma\eta\etap$ including all possible intermediate states is given
at the 90\% C.L.

The corresponding upper limits of the product branching fractions
$\BR(\pspp\to \gamma\xs)\cdot \BR(\xs\to \eta\etap)$ and
$\BR(\pspp\to \gamma\xs)\cdot \BR(\xs\to \ppjpsi)$ are calculated using
\begin{equation}\label{equ:42}
\begin{aligned}
 &\BR(\pspp\to\gamma \xs)\cdot\BR(\xs\to\eta\etap~(\ppjpsi)) \\
 &\quad\quad\quad\quad\quad\quad < \frac{N_{\eta\etap~(\ppjpsi)}^{\rm up}}
 {\lum\sigma W_{\eta\etap~(\ppjpsi)}}, \\
\end{aligned}
\end{equation}
where $N_{\eta\etap~(\ppjpsi)}^{\rm up}$ is the upper limit of the signal yield at the 90\% C.L.,
$\lum$ is the integrated luminosity of the used data sample, $\sigma = (7.15\pm
0.27\pm0.27)$ nb, is the observed
cross section of $\EE\to \pspp$ at 3.773 GeV~\cite{sys3, sys4}, $W_{\eta\etap}=\eff_{0}\BR(\eta\to\gamma\gamma)\BR(\etap\to\gamma\pp)$,
$W_{\ppjpsi}=\eff_{0}^{\ell\ell}\BR(\jpsi\to\LL)$.
$\BR(\eta\to\gamma\gamma)$, $\BR(\etap\to\gamma\pp)$, and $\BR(\jpsi\to\LL)$ are
taken from the PDG~\cite{int22}.
$\eff_{0}$ and $\eff_{0}^{\ell\ell}$ are the detection efficiencies based on MC simulation for $\xs\to\eta\etap$ and $\xs\to\ppjpsi$ modes, respectively.

\section{SYSTEMATIC UNCERTAINTY}\label{sec:sys unc}
The systematic uncertainties are classified into two categories: the multiplicative ones and the additive ones.  The multiplicative systematic uncertainties are listed in
Table~\ref{total sys err} and discussed in detail below.

The sources of multiplicative systematic uncertainties include the luminosity measurements,
the cross section of $\EE\to \pspp$, the data-MC differences of tracking efficiency and
photon reconstruction efficiency, the branching fractions taken from the PDG~\cite{int22},
the kinematic fit, the mass windows for $\etap$, $\eta\etap$, and $\jpsi$ candidates,
the requirement of $\cos\theta_{\gamma}$, and other selection criteria.

The integrated luminosity of the used data sample is measured using large angle Bhabha scattering
events, with an uncertainty of $0.5\%$~\cite{data1, data2}. The observed cross section
for $\EE\to \pspp$ is obtained by weighting the two independent measurements of this
cross section~\cite{sys3,sys4}, with an uncertainty of $5.3\%$.
The difference in the tracking efficiencies between data and MC simulation is estimated
to be 1\% per track~\cite{sys5,sys6}. The uncertainty due to the photon reconstruction efficiency
is studied in Refs.~\cite{sys74,sys75} and it is determined
to be $1\%$ per photon. The uncertainties from the branching fractions of
$\eta\to \gamma\gamma$, $\etap\to \gamma\pp$ and $\jpsi\to \LL$ are taken from
the PDG~\cite{int22}.

For the uncertainty caused by the kinematic fit to the
charged decay modes, we correct the track helix parameters
in the MC simulation so that the MC simulation can better
describe the momentum spectra of the data~\cite{sys9}.
In this analysis, we use the efficiency after the helix correction for the nominal results.
The difference in the MC efficiencies before and after performing the correction is
taken as the systematic uncertainty.

The efficiencies of many selection criteria used in this analysis are estimated
with control samples and compared with corresponding
MC simulations. The efficiency obtained from
MC simulation is corrected according to the data-MC difference in efficiencies obtained with
the control samples. The correction factor $f^{v}$ is defined as
 \begin{equation}\label{equ:1}
   f^{v}= \eff_{\rm MC}^{v}/\eff_{\rm data}^{v},
 \end{equation}
with
 \begin{equation}\label{equ:2}
   \eff^{v}_{\rm data(MC)}= N^{v}_{\rm data(MC)}/M^{v}_{\rm data(MC)},
  \end{equation}
where the subscript ``MC'' represents MC simulation and the subscript ``data'' represents the data sample,
$N^{v}_{\rm data(MC)}$ is the number of events in the signal region of a selection
criterion $v$, and $M^{v}_{\rm data(MC)}$ is the number of events in the full
range of $v$.

The uncertainty of $\eff^{v}_{\rm data(MC)}$ is
 \begin{equation}\label{equ:3}
 \sigma_{\eff^{v}_{\rm data(MC)}}= \sqrt{\frac{\eff^{v}_{\rm data(MC)}(1-\eff^{v}_{\rm data(MC)})}
                                              {M^{v}_{\rm data(MC)}}},
\end{equation}
since the data and MC simulation are independent, so the uncertainty of $f^{v}$ is
 \begin{equation}\label{equ:4}
 \frac{\sigma^{2}_{f^{v}}}{{f^{v}}^{2}}=\frac{\sigma^{2}_{\eff^{v}_{\rm data}}}
 {{\eff^{v}_{\rm data}}^{2}}+\frac{\sigma^{2}_{\eff^{v}_{\rm MC}}}{{\eff^{v}_{\rm MC}}^{2}}.
 \end{equation}

For $f^{v}\pm\sigma_{f^{v}}$, we define $\Delta f^{v} = |f^{v} - 1 |$, where $\Delta f^{v}$ and $\sigma_{f^{v}}$ are the deviation of $f^{v}$ from 1 and the uncertainty of $f^{v}$, respectively.
If $|\frac{\Delta f^{v}}{\sigma_{f^{v}}}| \leq 1.0$, no correction will
be applied ($f^{v}_{\rm cor}\equiv 1$) and $|\Delta f^{v}|+|\sigma_{f^{v}}| $ will be taken as the systematic uncertainty $\sigma_{f}$;
while if $|\frac{\Delta f^{v}}{\sigma_{f^{v}}}| > 1.0$, the MC efficiency $\eff$ will be corrected
as $\eff / f^{v} $, and $\sigma_{f^{v}}$ will be taken as the systematic uncertainty.
The systematic uncertainties due to the $\eta'$ and $\eta\eta'$ mass windows
are estimated by the control sample of $\psip\to\gamma\chi_{c0}$ (with $\chi_{c0}\to\eta\etap$). The systematic uncertainties due to the $\jpsi$ mass window and the $\cos\theta_{\gamma}$ requirement
are estimated by the control sample of $\EE\to\gamma_{\rm ISR}\psip$ (with $\psip\to\ppjpsi$).
The $f^{v}$ and $\sigma_{f}$  for the systematic uncertainties due to the $\eta'$, $\eta\eta'$, $\jpsi$ mass windows, and the $\cos\theta_{\gamma}$ requirement are shown in Table~\ref{totalsys}, where  $f^{v}_{\rm cor}$ and $\sigma_{f}$ are the final correction factor and systematic uncertainty, respectively.

\begin{table}[!htbp]
\caption{\label{totalsys}The $f^{v}\pm\sigma_{f^{v}}$, $f^{v}_{\rm cor}$ and $\sigma_{f}$ for the $\eta'$, $\eta\eta'$, $\jpsi$ mass windows, and the $\cos\theta_{\gamma}$ requirement. $f^{v}\pm\sigma_{f^{v}}$ is the efficiency correction factor. According to the definition in Sec.~\ref{sec:sys unc}, $f^{v}_{\rm cor}$ is the final efficiency correction factor and $\sigma_{f}$ is the systematic uncertainty of $f^{v}_{\rm cor}$.}
\begin{tabular}{cccccc}
    \hline \hline
   Sources  &  &$f^{v}\pm\sigma_{f^{v}}$ &$f^{v}_{\rm cor}$   &$\sigma_{f}$  \\
   \hline
   $\eta'$ mass window  &  &$1.0539\pm 0.0224$  &1.0539  &0.0224   \\
   $\eta\eta'$ mass window &   &$1.0076\pm 0.0199$ &1.0000 &0.0275  \\
   \multirow{2}{*}{$\jpsi$ mass window} &$\EE$ &$1.0039\pm 0.0025$ &1.0039  &0.0025 \\
                                         &$\MM$ &$1.0056\pm 0.0021$ &1.0056   &0.0021\\
    \multirow{2}{*}{$\cos\theta_{\gamma}$ requirement}   &$\EE$ &$1.0051\pm 0.0083$ &1.0000  &0.0134\\
                                                          &$\MM$  &$1.0007\pm 0.0069$ &1.0000 &0.0076\\
    \hline
     \hline
\end{tabular}
\end{table}

The efficiencies for other selection criteria, including the opening angle requirement, the event
start time determination and the $\chi_{\pi^{+}}$, $\chi_{\pi^{-}}$ requirements,
are higher than 99\%, and their total systematic uncertainties are safely assigned to be 1.0\%~\cite{sys10, sys11}.

Table~\ref{total sys err} summarizes all the multiplicative systematic uncertainties of
the two processes. The overall multiplicative systematic uncertainties are
obtained by adding all systematic uncertainties in quadrature
assuming they are independent.

The additive systematic uncertainties contain those from the parametrization of
the signal and background shapes in fitting $M(\ppjpsi)$ distributions
in the decay channel $\pspp\to \gamma\ppjpsi$.
The uncertainty due to the signal shape is derived from the difference in the mass
resolutions between data and MC simulation.
To estimate this uncertainty, an alternative fit is performed where the $\psip$ signal from $\EE\to \gamma_{\rm ISR}\psip$ (with $\psip\to \ppjpsi$) is
modeled with the MC shape of the $\psip$ convolved with a Gaussian resolution function.
The parameters (mean $m$ and standard deviation $\sigma$) of the Gaussian function
are determined to be $m = (-0.23\pm 0.03$)~MeV/$c^2$, $\sigma = (0.79\pm 0.05)$~MeV/$c^2$
for $\jpsi\to \EE$ and $m = (-0.13\pm 0.02)$~MeV/$c^2$, $\sigma = (0.80\pm 0.04)$~MeV/$c^2$
for $\jpsi \to \MM$.
We convolve the MC-determined $\xs$ shape with
the Gaussian smearing functions whose parameters are determined above, and refit the data.
The uncertainty associated with the background shape is estimated by changing from a first order Chebyshev function to a second order one.
Then the most conservative upper limit from these alternative fits is chosen as the upper limit with the additive systematic uncertainties.

\begin{table}[!htbp]
\caption{\label{total sys err} The multiplicative systematic uncertainties (in units of \%) in  the branching fraction measurements for $\pspp\to\gamma\xs\to\gamma\eta\etap$ and~$\pspp\to\gamma\xs\to\gamma\ppjpsi$.}
\begin{tabular}{cccc}
    \hline \hline

    \multirow{2}{*}{Sources}      &{$\gamma\eta\etap$} &\multicolumn{2}{c}{$\gamma\ppjpsi$}  \\
               &               &\quad$ee$         &$\mu\mu$            \\
   \hline
    Integrated luminosity                   &0.5      &\quad0.5    &0.5    \\
    Total cross section of $\pspp$     &5.3       &\quad5.3     &5.3     \\
    Tracking efficiency                     &2.0       &\quad4.0     &4.0     \\
    Photon reconstruction                   &4.0       &\quad1.0     &1.0     \\
    Kinematic fit                           &2.9       &\quad2.9     &1.9      \\
    Branching fraction of $\jpsi$           &-         &\quad0.6    &0.6     \\
    Branching fraction of $\eta$            &0.5      &\quad-       &-         \\
    Branching fraction of $\etap$           &1.4       &\quad-       &-         \\
    $\jpsi$ mass window                     &-         &\quad0.3    &0.2     \\
     $\etap$ mass window       &2.2       &\quad-       &-         \\
     $\eta\etap$ mass window  &2.8       &\quad-       &-          \\
    $\cos\theta_{\gamma}$ requirement               &-         &\quad 1.3     &0.8     \\
    Other requirements                               &-         &\quad 1.0     &1.0      \\
    \hline
    Total                                   &8.5       &\quad7.5    &7.1         \\
    \hline \hline
\end{tabular}
\end{table}

\section{UPPER LIMIT OF BRANCHING FRACTION}
Based on the Bayesian method~\cite{upp1}, the upper limit of the product branching fraction for $\pspp\to\gamma \xs$,
$\xs\to\eta\etap$ at the 90\% C.L. is calculated with the systematic uncertainty taken into account through the distribution of $M(\eta\eta')$ as shown in Fig.~\ref{pic:Metaetap-etap-data}(b).
To obtain the upper limit of the product branching fraction, the likelihood function is constructed to calculate the signal yield
at the 90\% C.L. assuming that the numbers of observed events in signal region ($N_{s+b}$) and
events in sideband region ($N_{b}$) obey a Poisson distribution ($Pois(N_{s+b};\mu+b)$, and $Pois(N_{b};b)$), and the efficiency ($\eff$) obeys a Gaussian distribution ($Gaus(\eff;\eff_{0},\delta^{\eta\etap}_{\eff})$).
The $\mu$ is the expected signal yield in the signal region, $b$ is the expected background yield
in the sideband region,
$\eff_{0}$ is the corrected efficiency,
$\delta^{\eta\etap}_{\eff}$ is the absolute multiplicative systematic uncertainty
for $\pspp\to \gamma\eta\etap$, $\delta^{\eta\etap}_{\eff}= \eff\times\delta^{\eta\etap}_{\rm sys}$,
where $\delta^{\eta\etap}_{\rm sys}$ is the relative total multiplicative systematic uncertainty
for $\pspp\to \gamma\eta\etap$.

Since there is almost no event left in the sideband region, we fix $N_{b}$ and $b$ to 0 in the likelihood
function to get a more conservative upper limit. The likelihood function is defined as
 \begin{equation}\label{equ:41}
 \begin{aligned}
        &{\rm L}(\mu) = \int_{0}^{1} {\rm L^{\prime}}(\frac{\eff}{\eff_{0}}\mu)\frac{1}{\sqrt{2\pi}\delta^{\eta\etap}_{\eff}}
        e^{-\frac{{(\eff-\eff_{0})}^{2}}{2{\delta^{\eta\etap}_{\eff}}^{2}}}d\eff, \\
        &\quad\quad\quad\quad{{\rm L^{\prime}}(\mu) = \frac{\mu^{N_{s}}}{N_{s}!}e^{-\mu} }.
 \end{aligned}
 \end{equation}

The signal yield ($N^{\rm up}_{\eta\etap}$) at the 90\% C.L. is determined as
 \begin{equation}\label{equ:10}
   \int_{0}^{N_{\eta\etap}^{\rm up}}{\rm L}(\mu)d\mu = 0.9\int_{0}^{\infty}{\rm L}(\mu)d\mu,
 \end{equation}
and the upper limit of the product branching fraction
$\BR(\pspp\to\gamma \xs)\cdot\BR(\xs\to\eta\etap)$ is calculated using Eq.~(\ref{equ:42}).
The obtained results are shown in Table~\ref{uplimit-Metaetap}.

\begin{table*}[htbp]
\renewcommand\arraystretch{1.5} 
\caption{\label{uplimit-Metaetap} The upper limit of the product branching fraction
$\BR(\pspp\to\gamma \xs) \cdot \BR(\xs\to\eta\etap)$ and
$\BR(\pspp\to\gamma \xs) \cdot \BR(\xs\to\ppjpsi)$ at the 90\% C.L.}
  \centering
\begin{tabular}{cccccccccc}
\hline
\hline
    $M(\xs)$ (MeV/$c^2$)   &  &$3710$  &$3715$    &$3720$   &$3725$    &$3730$  &$3733$ &$3735$   &$3740$   \\
 \hline
  \multirow{2}{*}{$\pspp\to\gamma\eta\etap$} &$\eff_{0}$ (\%)     &10.56 &10.50   &10.14     &9.77    &9.31   &8.99  &8.70 &8.15 \\

        &$\BR^{\rm up}(\times 10^{-6})$   &8.9  &9.0  &9.3 &9.7 &10 &18 &18 &19  \\
\hline
 \multirow{3}{*}{$\pspp\to\gamma\ppjpsi$}   &$\eff^{ee}_{0}$ (\%)    &15.68  &15.59 &15.88 &15.50 &15.38 &15.32 &14.83 &13.76  \\
      &$\eff^{\mu\mu}_{0}$ (\%)    &24.10 &23.97 &24.02 &23.93 &23.67 &23.54 &23.16 &21.80  \\
     &$\BR^{\rm up}(\times 10^{-5})$   &2.2  &1.2  &1.8 &3.0 &0.86 &1.0 &1.3 &3.4  \\
\hline
\hline
\end{tabular}
\end{table*}

The simultaneous unbinned maximum likelihood fit is performed on the $M(\ppjpsi)$ distributions
of $\jpsi\to \EE$ and $\jpsi\to\MM$ modes for $\pspp\to \gamma\ppjpsi$, and the likelihood distribution is obtained by varying the signal yield in the fit.
The normalized likelihood distribution is smeared with a Gaussian function, whose mean is
set as $\eff^{\ell\ell}_{0}$, and the standard deviation is
$\delta^{\ppjpsi}_{\eff} = \eff\times\delta^{\ppjpsi}_{\rm sys}$, where $\delta^{\ppjpsi}_{\eff}$ and $\delta^{\ppjpsi}_{\rm sys}$ are the absolute multiplicative systematic uncertainty and the relative total multiplicative systematic uncertainty for $\pspp\to \gamma\ppjpsi$, respectively.
The likelihood function is defined as Eq.~(\ref{equ:41}) and the signal yield ($N^{\rm up}_{\ppjpsi}$) at the 90\% C.L. is determined using Eq.~(\ref{equ:10}).
The upper limit of the product branching fraction $\BR(\pspp\to\gamma \xs)\cdot \BR(\xs\to\ppjpsi)$
is calculated using Eq.~(\ref{equ:42}) and the obtained results are shown in Table~\ref{uplimit-Metaetap}.

\begin{figure}[htbp]
  \centering
  \includegraphics[width=0.45\textwidth]{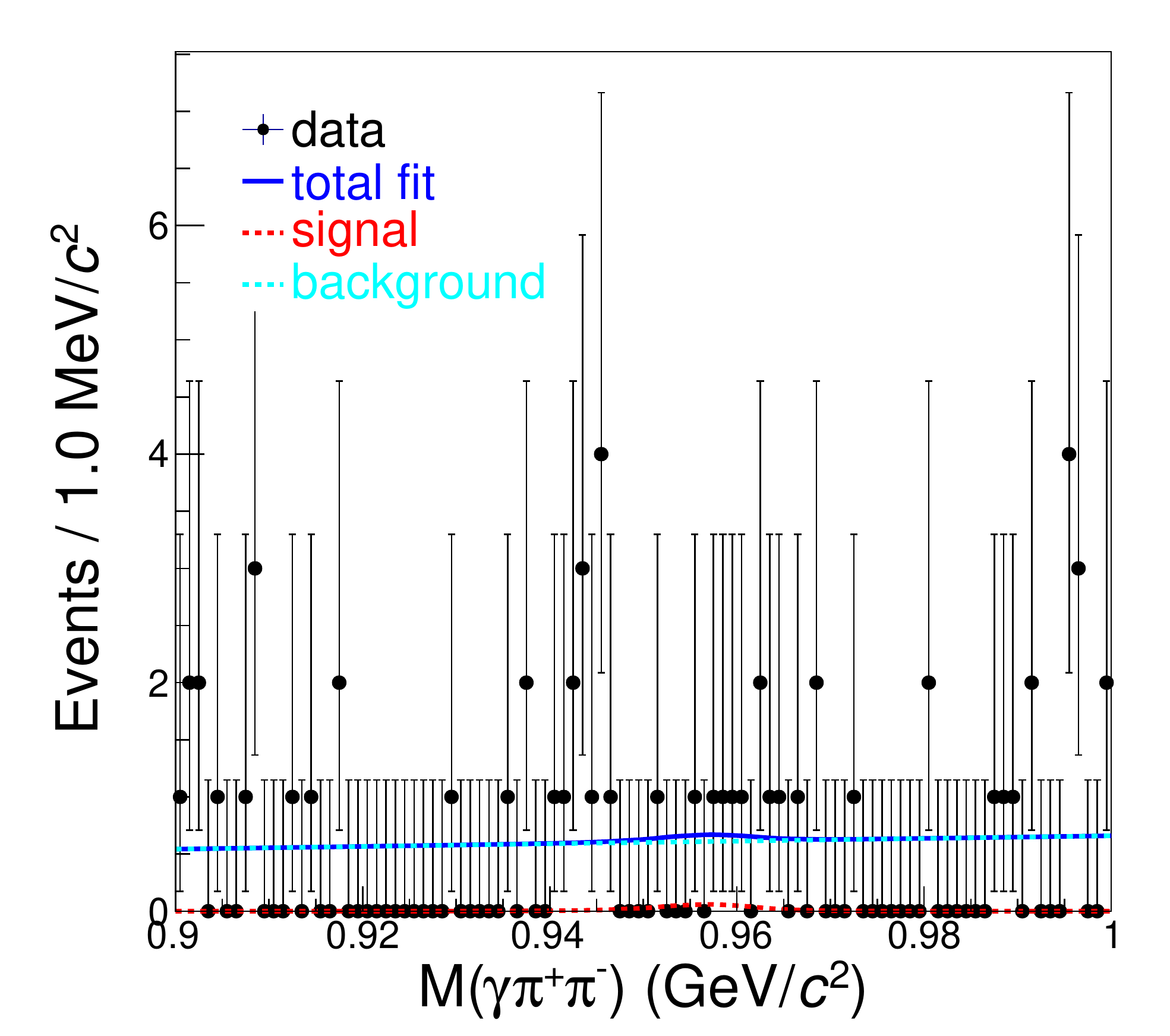}
\caption{The unbinned maximum likelihood fit on the M(\gampipi) distribution. Black dots with error bars represent data sample, red dashed line represent signal MC sample, background contributions are described by the green dashed line and blue solid line represents the total fit.}
\label{pic:fitetap}
\end{figure}

In addition, we also search for $\pspp\to\gamma\eta\etap$. The unbinned maximum likelihood fit is performed on the $M(\gamma\pp)$ distribution, which is shown in Fig.~\ref{pic:fitetap}. Since no obvious $\eta'$ signals are observed, the likelihood distribution is obtained by varying the signal yield in the fit.
The normalized likelihood distribution is smeared with a Gaussian function,
and the likelihood function is defined as Eq.~(\ref{equ:41}).
The systematic uncertainty is the same as that of $\xs\to\eta\etap$, except for excluding the systematic uncertainty due to the $\eta\eta'$ mass window.
The signal yield ($N^{\rm up}_{\gamma\eta\etap}$) at the 90\% C.L. is determined
using Eq.~(\ref{equ:10}).
The upper limit of the branching fraction for $\pspp\to\gamma\eta\etap$ is calculated as
\begin{equation}\label{equ:12}
\begin{aligned}
 &\quad\quad \quad\BR(\pspp\to\gamma\eta\etap) < \\
 &\frac{N_{\gamma\eta\etap}^{\rm up}}{\lum\sigma\eff^{\gamma\eta\etap}_{0}\BR(\eta\to\gamma\gamma)\BR(\etap\to\gamma\pp)},
\end{aligned}
\end{equation}
where $\eff^{\gamma\eta\etap}_{0}$ is the efficiency obtained by MC simulation,
which is 13.02\%. The upper limit of $\BR(\pspp\to\gamma\eta\etap)$ is determined to be
$4.8\times 10^{-5}$ at the 90\% C.L.

\section{SUMMARY and discussions}
Using 2.93~fb$^{-1}$ of data taken at $\sqrt s =3.773$ GeV accumulated in the BESIII experiment, we search for a scalar partner of the $\xx$, denoted as $\xs$, via $\pspp\to\gamma\eta\etap$ and $\pspp\to\gamma\ppjpsi$ for the first time.
Since no signal events are observed, the upper limits of the product branching fraction for these
two decay channels are determined at the 90\% C.L. and listed in Table~\ref{uplimit-Metaetap}.
We also give the upper limit of the branching fraction for $\pspp\to\gamma\eta\etap$
including all possible intermediate states for the first time, which is $4.8\times 10^{-5}$.

According to the theoretical prediction in Ref.~\cite{int21} that we have discussed in the introduction, the product of the branching fraction $\BR(\pspp\to\gamma \xs)\cdot\BR(\xs\to\eta\etap)$ = $1.08\times 10^{-5}$. The upper limits for this variable that we obtain are from 0.9 to 1.9 $(\times 10^{-5})$, for the $\xs$ mass ranging from 3710 to 3740 MeV/$c^{2}$.
The results are lower than the theoretical expectations for most of the possible $\xs$ masses, which disfavor the theoretical prediction that $\eta\etap$ is the dominant channel for the $\xs$ decay.

$\xs\to\ppjpsi$ has also been searched for, but
no significant $\xs$ signals are found.
The upper limits of the product branching fraction are calculated to be from 0.9 to 3.4 $(\times 10^{-5})$ at the 90\% C.L., for the $\xs$ mass ranging from 3710 to 3740 MeV/$c^{2}$.
As the spin partner of the $\xx$, the $\xs\to\ppjpsi$ might have large isospin violation
similar to the $\xx\to\ppjpsi$~\cite{int24}. This result can provide a constraint for the theoretical calculation of $\xs$ .

\section{Acknowledgement}
The BESIII Collaboration thanks the staff of BEPCII and the IHEP computing center for their strong support. This work is supported in part by National Key R\&D Program of China under Contracts Nos. 2020YFA0406300, 2020YFA0406400; National Natural Science Foundation of China (NSFC) under Contracts Nos. 11635010, 11735014, 11835012, 11935015, 11935016, 11935018, 11961141012, 12022510, 12025502, 12035009, 12035013, 12061131003, 12192260, 12192261, 12192262, 12192263, 12192264, 12192265, 12221005, 12225509, 12235017; the Chinese Academy of Sciences (CAS) Large-Scale Scientific Facility Program; the CAS Center for Excellence in Particle Physics (CCEPP); CAS Key Research Program of Frontier Sciences under Contracts Nos. QYZDJ-SSW-SLH003, QYZDJ-SSW-SLH040; 100 Talents Program of CAS; The Institute of Nuclear and Particle Physics (INPAC) and Shanghai Key Laboratory for Particle Physics and Cosmology; ERC under Contract No. 758462; European Union's Horizon 2020 research and innovation programme under Marie Sklodowska-Curie grant agreement under Contract No. 894790; German Research Foundation DFG under Contracts Nos. 443159800, 455635585, Collaborative Research Center CRC 1044, FOR5327, GRK 2149; Istituto Nazionale di Fisica Nucleare, Italy; Ministry of Development of Turkey under Contract No. DPT2006K-120470; National Research Foundation of Korea under Contract No. NRF-2022R1A2C1092335; National Science and Technology fund of Mongolia; National Science Research and Innovation Fund (NSRF) via the Program Management Unit for Human Resources \& Institutional Development, Research and Innovation of Thailand under Contract No. B16F640076; Polish National Science Centre under Contract No. 2019/35/O/ST2/02907; The Swedish Research Council; U. S. Department of Energy under Contract No. DE-FG02-05ER41374.

\nocite{*}

\end{document}